\documentclass[twocolumn,english]{article}
\usepackage[T1]{fontenc}
\usepackage[latin9]{inputenc}
\usepackage{geometry}
\usepackage{graphicx}

\providecommand{\tabularnewline}{\\}

\makeatother

\usepackage{babel}

\usepackage{babel}

\begin{document}
\textbf{\large Comment on {}``Three Extra Mirror or Sequential Families:
Case for a Heavy Higgs Boson and Inert Doublet\textquotedblright{}}{\large {} }{\large \par}

{\small \medskip{}
 }{\small \par}

M. Sahin$^{1}$, S. Sultansoy$^{1,2}$, G. Unel$^{3}$

{\small \medskip{}
 $^{1}$TOBB University of Economics and Technology, Ankara, Turkey
}\\
 {\small $^{2}$ANAS Institute of Physics, Baku, Azerbaijan }\\
 {\small $^{3}$University of California at Irvine, Irvine, CA,
USA\medskip{}
 }{\small \par}

In a recent Letter \cite{Martinez}, Martinez \emph{et al.} have argued
that the existence of extra three families is still perfectly in accord
with the standard model (SM), as long as an additional Higgs doublet
is also present. Here we claim that the mentioned accordance can be
obtained with less reliance on BSM physics like additional Higgs doublets,
but benefiting from Majorana neutrinos is sufficient. For the remainder
of the discussion we will denote SM with $n$ families as SM$n$ ($n$
= 3, 4, 5, 6).

Let us remember that while first papers on the SM did not contain
right-handed components of the neutrinos, their inclusion is necessary
according quark-lepton symmetry ($\nu_{R}$ is counterpart of $u_{R}$).
Moreover, if there are no any arguments (like exact conservation of
lepton charge) forbidding Majorana mass terms for $\nu_{R}$, these
terms should be taken into account. Of course, this leads to some
complication (such as the $2n\times2n$ mass matrix in neutrino sector
instead of an $n\times n$ one) but, simultaneously, provides small
masses for neutrinos from the first three families via see-saw mechanism.
Therefore, Majorana nature for the SM neutrinos seems more natural
than Dirac one.

The widely accepted method for measuring the compatibility of a model
with additional fermions and the electroweak (EW) precision data is
utilization of the oblique parameters \cite{stu-def}. The rather
tedious calculation of $S$ and $T$ parameters is usually computerized.
Among similar computer programs, OPUCEM \cite{opucem-pub} is the
only tool which can include Majorana neutrinos in the calculations.
In the simple SM4 case, the inclusion of Majorana neutrinos essentially
enlarges the parameter space, resulting in a large number of SM4 mass
points which are in better accordance with the data compared to SM3
\cite{Sahin}.

\begin{figure}
\begin{centering}
\includegraphics[scale=0.4]{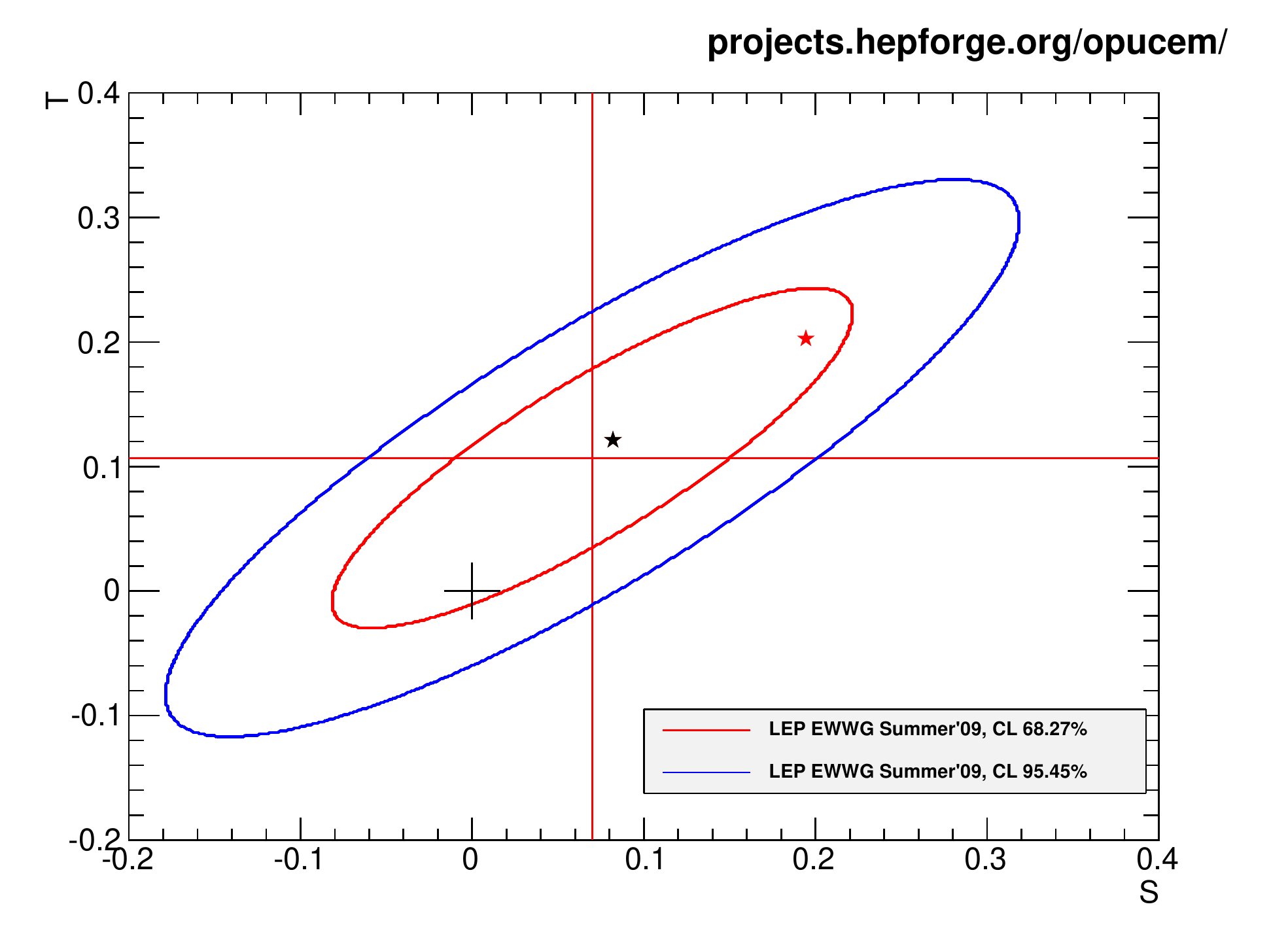} 
\par\end{centering}

\caption{Majorana type SM6 points with low (115 GeV, dark star) and high (600
GeV, light star) Higgs masses as compared to EW precision data error
ellipses and the SM3 prediction with $m_{h}=115$ GeV and $m_{t}=173$
GeV (cross). \label{fig:SM6-data-points} }

\end{figure}

\begin{table}[!h]
\caption{Two Majorana type SM6 mass points (mass in GeV). First point (P1)
with $m_{h}$=115GeV and $|\sin\theta_{34}|=0.08$, leading to $S$=0.082
and $T$=0.121 is on the left and the second point (P2) with $m_{h}$=600GeV
and $|\sin\theta_{34}|=0.06$, leading to $S$=0.194 and $T$=0.203
is on the right .\label{tab:Two-SM6-mass}}

\centering{}\begin{tabular}{|c|c|c|c||c|c|c|}
\hline 
 & 4th f.  & 5th f.  & 6th f.  & 4th f.  & 5th f.  & 6th f.\tabularnewline
\hline
\hline 
$m_{U}$  & 430  & 565  & 550  & 430  & 565  & 550\tabularnewline
\hline 
$m_{D}$  & 385  & 450  & 500  & 385  & 450  & 500\tabularnewline
\hline 
$m_{\nu}$  & 45  & 45  & 450  & 45  & 45  & 450\tabularnewline
\hline 
$m_{E}$  & 570  & 580  & 600  & 590  & 580  & 600\tabularnewline
\hline 
$m_{N}$  & 2600  & 2800  & 3200  & 2600  & 2800  & 3200\tabularnewline
\hline
\end{tabular}
\end{table}

Concerning SM5 and SM6 OPUCEM analysis show that in the Majorana case
there are a lot of mass points in 1$\sigma$ agreement with the precision
EW data. For illustration we present two SM6 points in S-T plane in
Fig. \ref{fig:SM6-data-points} with corresponding parameters presented
in Table \ref{tab:Two-SM6-mass} (where $\nu$ and $N$ denote light
and heavy Majorana neutrinos, respectively). It is seen that SM6 point
with $m_{h}=115$ GeV is closer to central value than SM3 point.

\begin{figure}
\includegraphics[scale=0.4]{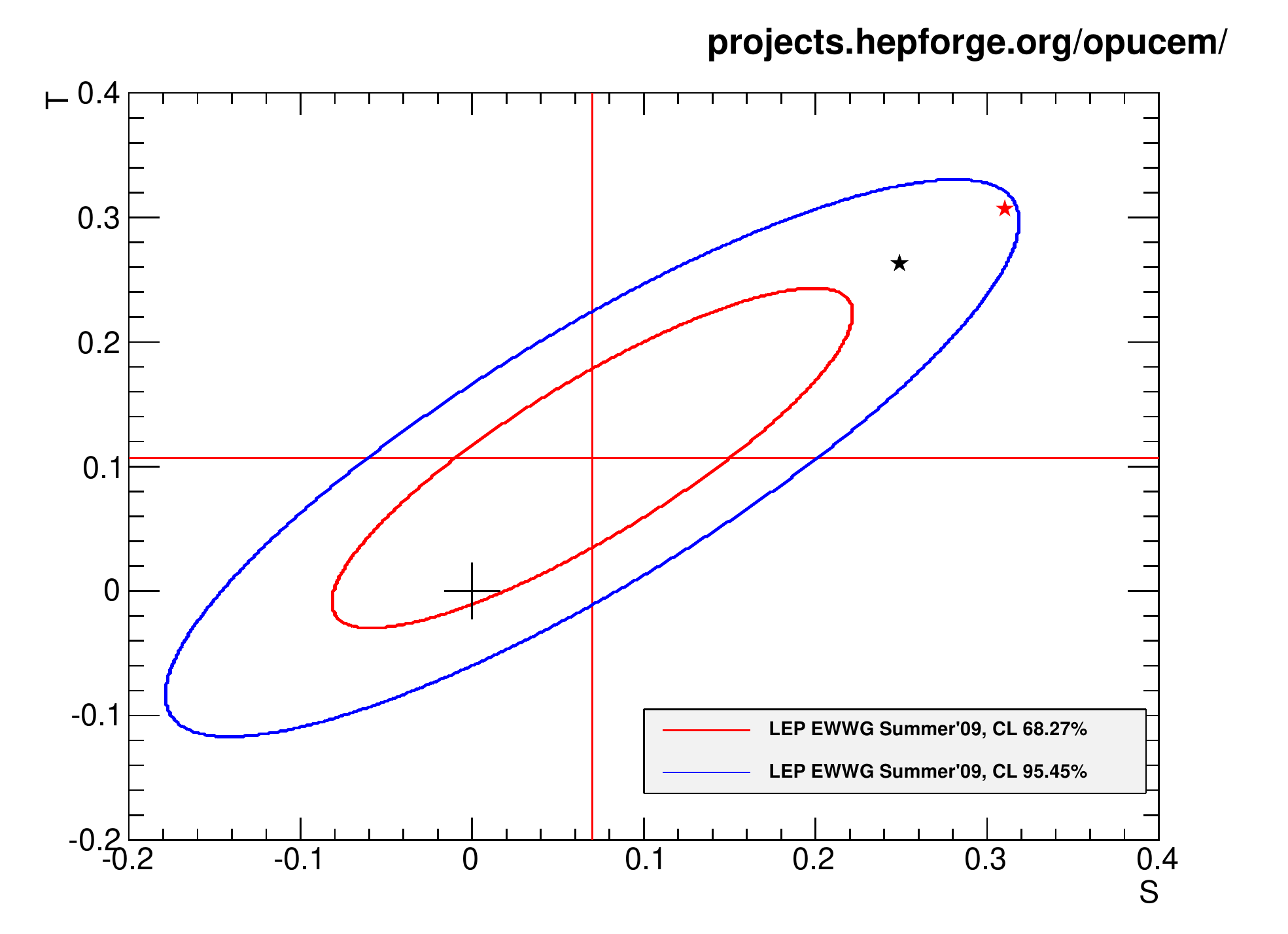}

\caption{Dirac type SM6 points with low (115 GeV, dark star) and high (600
GeV, light star) Higgs masses as compared to EW precision data error
ellipses and the SM3 prediction with $m_{h}=115$ GeV and $m_{t}=173$
GeV (cross).}

\end{figure}

\begin{table}
\caption{Two Dirac type SM6 mass points (mass in GeV). First point with $m_{h}=115$
GeV and $|sin\theta_{34}|=0.00$, leading to $S=0.25$ and $T=0.26$
is on the left and second point with $m_{h}=600$ GeV and $|sin\theta_{34}|=0.01$,
leading to $S=0.31$ and $T=0.31$ is on the right.}

\begin{tabular}{|c|c|c|c||c|c|c|}
\hline 
 & 4th f. & 5th f. & 6th f. & 4th f. & 5th f. & 6th f.\tabularnewline
\hline
\hline 
$m_{U}$  & $530$ & $550$ & $580$ & $580$ & $580$ & $590$\tabularnewline
\hline 
$m_{D}$  & $520$ & $550$ & $580$ & $580$ & $580$ & $590$\tabularnewline
\hline 
$m_{\nu}$  & $50$ & $50$ & $50$ & $50$ & $50$ & $50$\tabularnewline
\hline 
$m_{E}$  & $110$ & $130$ & $120$ & $110$ & $140$ & $180$\tabularnewline
\hline
\end{tabular}
\end{table}

Keeping in mind that sometimes Majorana neutrinos are interpreted
as BSM physics, OPUCEM results for Dirac neutrinos are presented in
Fig. 2 (corresponding parameters are presented in Table 2). It is
seen that even with Dirac neutrinos precision electroweak data allows
three additional SM families within $2\sigma$! Therefore, results
of \cite{Martinez} are neither fascinating nor surprising. 

Finally, let as mention that there are phenomenological arguments
against fifth (and certainly sixth) SM family (see, e.g. \cite{Sahin}
and references therein). Nevertheless, the fact that even SM6 is not
contradict to present EW data is quite intriguing.

\end{document}